\documentclass{jaa}
\usepackage{amsfonts}
\usepackage{amsmath}
\usepackage{amssymb}
\usepackage{amsbsy}
\usepackage[british]{babel}
\usepackage{latexsym}
\usepackage{mathtools}
\usepackage{setspace}
\usepackage{caption}
\usepackage{color}   
\usepackage{epsfig}
\usepackage{graphics}
\usepackage{graphicx}
\usepackage{xmpmulti}
\usepackage{aas_macros} 
\usepackage{astrobib} 
\def\i{\item}

\newcommand{\bed}{\begin{displaymath}}
\newcommand{\eed}{\end{displaymath}}
\newcommand{\bei}{\begin{itemize}}
\newcommand{\eei}{\end{itemize}}
\newcommand{\bef}{\begin{figure}}
\newcommand{\eef}{\end{figure}}
\newcommand{\ben}{\begin{enumerate}}
\newcommand{\een}{\end{enumerate}}
\newcommand{\beq}{\begin{equation}}
\newcommand{\eeq}{\end{equation}}
\newcommand{\ber}{\begin{eqnarray}}
\newcommand{\eer}{\end{eqnarray}}

\newcommand{\bb}{\bf B}

\newcommand{\vb}{\bf v}

\newcommand{\pa}{\partial}

\newcommand{\gcc}{\mbox{${\rm g} \, {\rm cm}^{-3}$}}

\newcommand{\mdot}{\mbox{$\dot{M}$}}
\newcommand{\msun}{\mbox{{\rm M}$_{\odot}$}}

\newcommand{\pdot}{\mbox{$\dot P$}}

\newcommand{\dmdt}{{\mbox{{\rm M}$_{\odot}$}} {\rm yr}$^{-1}$}

\newcommand{\lsim}{\raisebox{-0.3ex}{\mbox{$\stackrel{<}{_\sim} \,$}}}
\newcommand{\gsim}{\raisebox{-0.3ex}{\mbox{$\stackrel{>}{_\sim} \,$}}}

\begin{document}

\title{Magnetic Fields of Neutron Stars}

\author{Sushan Konar\textsuperscript{*}}
\affilOne{NCRA-TIFR, Pune, 411007, India}


\twocolumn[{

\maketitle

\corres{sushan@ncra.tifr.res.in}


\begin{abstract}
  This  article  briefly  reviews  our current  understanding  of  the
  evolution  of  magnetic fields  in  neutron  stars, which  basically
  defines  the evolutionary  pathways between  different observational
  classes of neutron  stars. The emphasis here is on  the evolution in
  binary  systems  and  the  newly  emergent  classes  of  millisecond
  pulsars.
\end{abstract}


}]





\section{Introduction}
\label{sec01}

Over the  decades, one of  the primary preoccupations of  neutron star
research  has been  to look  for a  unification scheme  connecting the
widely     different      observational     classes      (shown     in
Fig.\ref{f_menagerie}).  The magnetic field,  ranging from $10^8$~G in
millisecond pulsars to  $10^{15}$~G in magnetars, has  been central to
this theme as it plays an  important role in determining the evolution
of the spin, the radiative properties and the interaction of a neutron
star with its  surrounding medium.  Consequently, it  is the evolution
of  the magnetic  field  which links  different observational  classes
through unique evolutionary pathways \cite{bhatt95c,bhatt02,reise07b}.

The  nature  of field  evolution  depends  crucially on  the  material
properties  of the  region  where currents  supporting  the field  are
located. In  the solid  crust, the currents  are carried  by electrons
making Ohmic diffusion  and Hall drift the  predominant mechanisms for
field   evolution   \cite{goldr92}.     While   Ohmic   diffusion   is
intrinsically                                              dissipative
\cite{sang87,urpin92a,urpin93,geppe94,urpin95b,konar97a,konar99a,urpin08},
Hall drift is basically an  advection of the magnetic field.  Although
Hall drift is  not a dissipative process, it  can cause rearrangements
of   the   field   which    may   then   enhance   Ohmic   dissipation
\cite{holle02,rhein02,geppe02,reise07a,pons09,vigan13}.

On  the other  hand, in  the  core of  a neutron  star the  electrical
conductivities are  so large that  Hall drift and Ohmic  diffusion are
extremely  slow  to  have   any  effect  on  astrophysically  relevant
time-scales.   However, the  core  is a  fluid  mixture of  electrons,
protons and neutrons (and perhaps other more exotic particles) and the
magnetic flux can be transported and re-arranged by convective motions
for a  period immediately  after the  birth of  the neutron  star.  At
later times (when the temperature is low enough) the charged particles
and  the associated  magnetic  field  can drift  with  respect to  the
neutrons,    in    a     process    called    `ambipolar    diffusion'
\cite{goldr92,passa17}.   

However, protons  inside the core  of a  neutron star are  expected to
behave like  a type-II superconductor  with a upper critical  field of
$\sim 10^{16}$~G. In this case, magnetic field would exist in the form
of an  array of  Abrikosov fluxoids.  Recently,  the MHD  behaviour of
such superconductors  has been investigated  and it appears  that the
time-scales     of    field     evolution    are     extremely    long
\cite{glamp11a,grabe15,elfri16}. Therefore,  the processes responsible
for field evolution  can only be effective if  the currents supporting
the field are located (or are relocated in the course of evolution) in
the  crustal  region which  has  metal-like  transport properties.   A
number of mechanisms have been  proposed for the expulsion of magnetic
flux  from  the  core to  the  crust  \cite{musli85b,srini90b,konen00}
without reaching any consensus.

The evolution of the magnetic  field is also significantly affected by
external  processes   like  material  accretion.   For   example,  the
accretion  phase  connects  up   ordinary  radio  pulsars  with  their
millisecond counterparts through a  well defined evolutionary pathway.
The  rotation-powered millisecond  pulsars are  understood to  acquire
their  low magnetic  fields  and fast  spin  frequencies through  mass
accretion in LMXBs.  For this pathway to work, it has been argued that
a  physical model  of  field evolution  should  satisfy the  following
observational constraints  -- relatively  little magnetic  field decay
should take place in an  isolated radio pulsar, while accretion should
be able  to reduce  the surface  field strength  by several  orders of
magnitude \cite{bhatt02,konar13}.   On the  other hand,  to understand
the connection  between different  types of  isolated neutron  stars a
detailed  theory of  magneto-thermal evolution  has been  developed in
recent  years  \cite{pons09,kaspi10,vigan13}.  The  resulting  synergy
between  the  theoretical  models   and  the  observation  has  indeed
strengthened  the notion  of a  magnetic field  dominated evolutionary
link between different observational classes of neutron stars.


\section{The Neutron Star Menagerie}
\label{sec02}

\begin{figure*}
%
\begin{center}
\includegraphics[angle=-90,width=10.0cm]{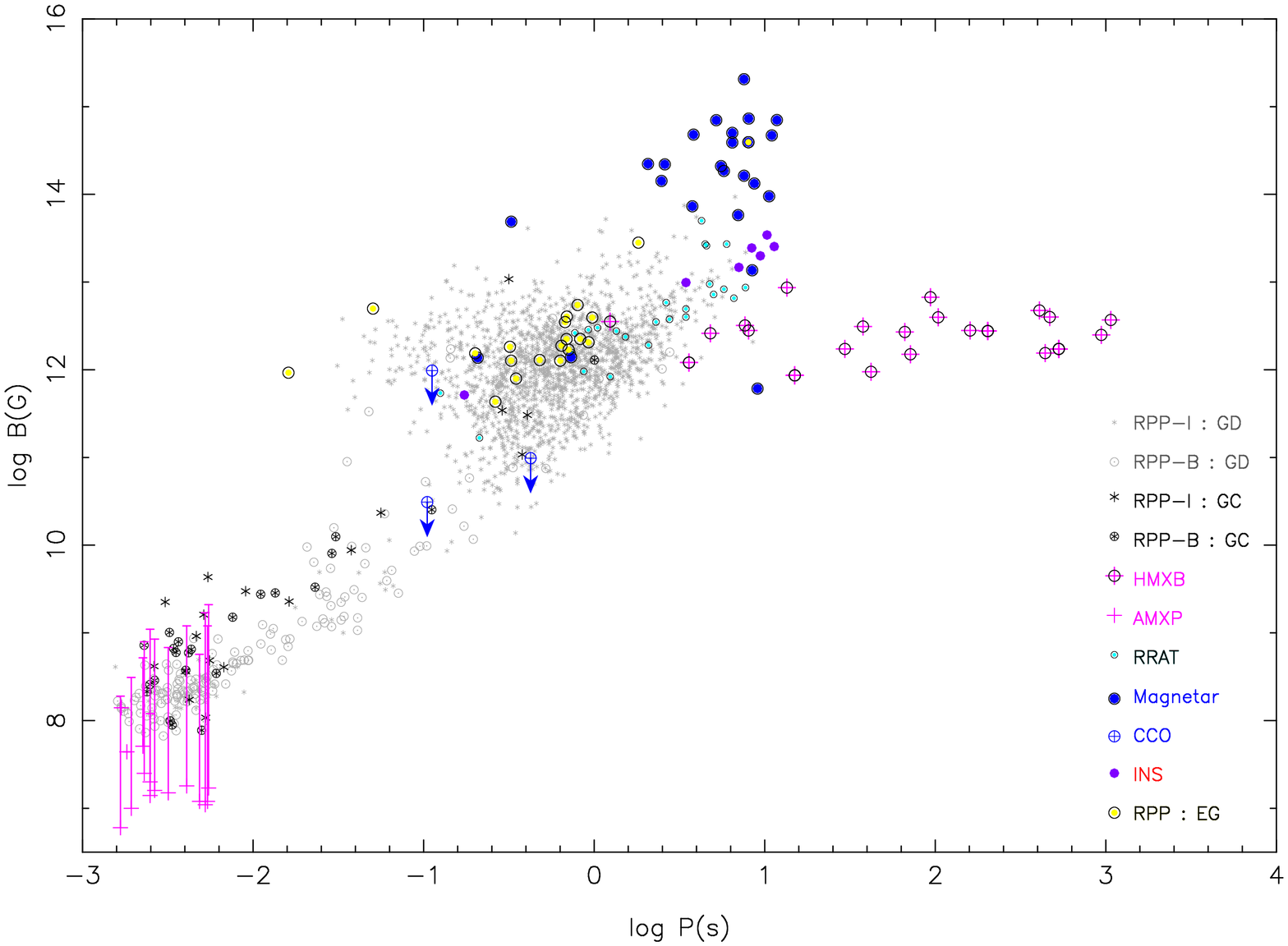}
\end{center}
%
\caption{The Menagerie : Different  observational classes  of neutron
  stars  in the P$_{\rm s}$--B$_{\rm s}$ plane.  \\
  {\bf \em Legends  :} I/B - isolated/binary, GC -
  globular cluster, GD - galactic disc, EG - extra-galactic objects. \\
  {\bf \em Data :} 
  RPP - {\tt http://www.atnf.csiro.au/research/pulsar/psrcat/},\\
RRAT - {\tt http://astro.phys.wvu.edu/rratalog/},\\
Magnetar - {\tt http://www.physics.mcgill.ca/~pulsar/magnetar/main.html},\\
AMXP - Patruno \& Watts (2012), Mukherjee et al. (2015);
HMXB - Caballero \& Wilms (2012),
INS - Haberl (2007), Kaplan \& van Kerkwijk (2009);
CCO - Halpern \& Gotthelf (2010), Ho (2013).
}
\label{f_menagerie}
\end{figure*}
\nocite{patru12c,mukhe15,cabal12,haber07,kapla09a,halpe10,ho13}

About 3000 neutron stars, with diverse characteristic properties, have
been observed  till date,  in different  parts of  the electromagnetic
spectrum.  Interestingly,  these  can  be classified  into  the  three
following  types, according  to  the nature  of  of energy  generation
processes in them \cite{kaspi10,konar13,konar16c}.

{\bf  Rotation Powered  Pulsars  (RPP)  :} Discovered  serendipitously
\cite{hewis68}  and  originally  known  as radio  pulsars,  these  are
powered by the loss of rotational energy due to magnetic braking.  The
basic observed quantities  of a pulsar are  the spin-period (P$_{\rm
  s}$) and  the period  derivative ($\pdot$),  and the  most important
derived  quantity is  the dipolar  component of  the surface  magnetic
field given by \cite{manch77},
\beq
B_{\rm s} \simeq 3.2 \times 10^{19}
\left(\frac{P_{\rm s}}{s}\right)^{\frac{1}{2}}
\left(\frac{\pdot}{ss^{-1}}\right)^{\frac{1}{2}}~{\rm G}\,.
\label{eq01}
\eeq
The RPPs again have three sub-classes.
\ben
   \i Primary  among the  RPPs are the  classical radio  pulsars (PSR)
   with P$_{\rm s} \sim 1$s and B$_{\rm s} \sim 10^{11.5} - 10^{13.5}$~G.
   \i Detection of the  1.5~ms pulsar B1937+21 \cite{backe82} heralded
   this  new genre  of millisecond  radio pulsars  (MSRP).  MSRPs  are
   usually  defined  as  the  ones  with   $P  \  \lsim  20  -  30$~ms
   \cite{lorim09}.  The definition is  not accurate as the requirement
   is to separate out a class of RPPs that have gone through different
   evolutionary  pathways involving  long-lived binary  systems and  a
   `recycling' accretion episode reducing both the spin-period and the
   magnetic field  \cite{tauri11}.  Fortunately though, the  ranges of
   P$_{\rm s}$ and B$_{\rm s}$ of  these MSRPs clearly place them in a
   nearly disjoint region  of the P$_{\rm s}$--B$_{\rm  s}$ plane from
   the normal radio pulsars (Fig.\ref{f_menagerie}).
   \i The  {\em black widow} (BW)  and the {\em redback}  (RB) pulsars
   actually  comprise a  special class  of binary  MSRPs. They  are so
   named  because  they  are  in   the  process  of  destroying  their
   companions through strong pulsar winds and are likely to be left as
   isolated millisecond pulsars \cite{kluzn88,phinn88}.
\een

{\bf Accretion Powered Pulsars (APP)  :} Powered by material accretion
from a  companion, these  are classified  as High-Mass  X-ray Binaries
(HMXB) or Low-Mass X-Ray Binaries (LMXB)  depending on the mass of the
donor star.
\ben
   \i Neutron stars in HMXBs with  B$_{\rm s} \sim 10^{12}$~G have $O$
   or  $B$ type  companions and  typically  show up  as X-ray  pulsars
   \cite{cabal12}.   Interaction of  the accreting  material with  the
   strong  magnetic  field of  the  neutron  star  shows up  as  broad
   absorption lines  known as Cyclotron Resonance  Scattering Features
   (CRSF)  in  the hard  X-ray  spectrum  of  HMXBs. Strength  of  the
   magnetic field can be estimated from CRSFs as follows -
      \beq
      E_{c} = 11.6  \mbox{keV}
      \times \frac{1}{1+z} \times \frac{B}{10^{12}}~\mbox{G};
      \eeq
      where  $E_{c}$   is  the   centroid  energy   and  $z$   is  the
      gravitational red-shift.
   \i Neutron stars in LMXBs, on  the other hand, have magnetic fields
   significantly weakened  (B$ \lsim  10^{11}$~G) through  an extended
   phase of accretion. Among all  the LMXBs harbouring neutron stars -
   the  accreting  millisecond  X-ray  pulsars ({\bf  AMXP})  and  the
   accreting millisecond  X-ray bursters  ({\bf AMXB}) are  of special
   interest as they are understood to be immediate precursors of MSRPs
   and are expected to go into a radio pulsar phase upon the cessation
   of  accretion.  The  magnetic  field in  these  systems is  usually
   obtained based upon  an estimate of the inner  truncation radius of
   the  accretion disc,  since at  that  radius the  disk pressure  is
   balanced by the pressure of the magnetic field.
\een
 
{\bf Internal Energy Powered Neutron  Stars (IENS) :} The mechanism of
energy generation is  not obvious for this class of  neutron stars and
is suspected to be related to some internal process, like the decay of
a  strong magnetic  field or  the  residual heat  from the  progenitor
supernova.
\ben
   \i {\em Magnetars} are thought to be young, isolated neutron stars,
   that  manifest themselves  as soft  gamma ray  repeaters (SGR)  and
   anomalous X-ray pulsars (AXP).  It is believed that the main energy
   source  of these  objects  is the  decay  of super-strong  magnetic
   fields (magnetar model) \cite{thomp96}.
   \i X-ray bright central compact  objects (CCO) are characterised by
   an  absence  both  of  associated  nebulae  and  exceptionally  low
   magnetic  fields   (B$  \sim  10^{10}$~G).   Likely,   a  phase  of
   hypercritical accretion immediately after  the birth of the neutron
   star  buries the  original field  to  deeper regions  of the  crust
   \cite{vigan12}.
   \i The seven  isolated neutron stars (INS), popularly  known as the
      {\em   Magnificent  Seven},   are  optically   faint  and   have
      blackbody-like  X-ray  spectra  ($T  \sim  10^6$~K).   They  are
      probably like ordinary pulsars (with the pulsar beam turned away
      from us) but a combination  of strong magnetic field and spatial
      proximity make them visible in X-rays.
\een

\section{Magnetic Field : Evolution in Binaries}
\label{sec03}

As mentioned  before the  evolution of  magnetic field  is one  of the
central  ingredients  in  understanding the  interconnections  between
different observational classes  of neutron stars. In  this section we
discuss  some   of  the  well-developed  formalisms   that  have  been
reasonably  successful in  explaining the  field evolution  in neutron
stars   engaged   in   active   mass  accretion,   as   indicated   by
observations\footnote{The  evolution  of  magnetic  field  in  neutron
  stars, in particular  the question of MSRP generation  in LMXBs, has
  been one of  the primary focus area of the  astrophysics group, RRI,
  under the leadership of G. Srinivasan}.

\setcounter{figure}{1}
\begin{figure*}
\includegraphics[width=160pt]{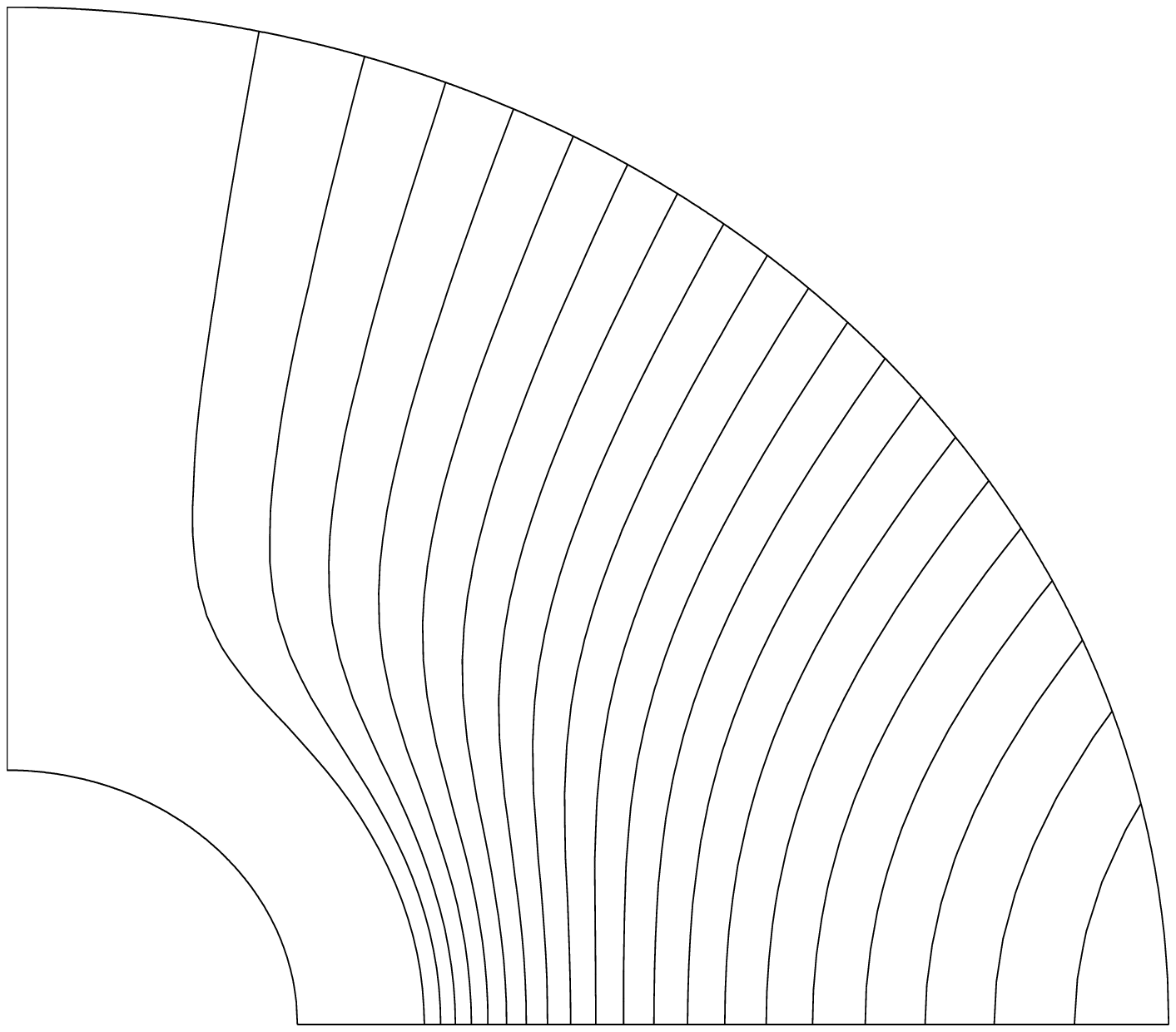}
\includegraphics[width=160pt]{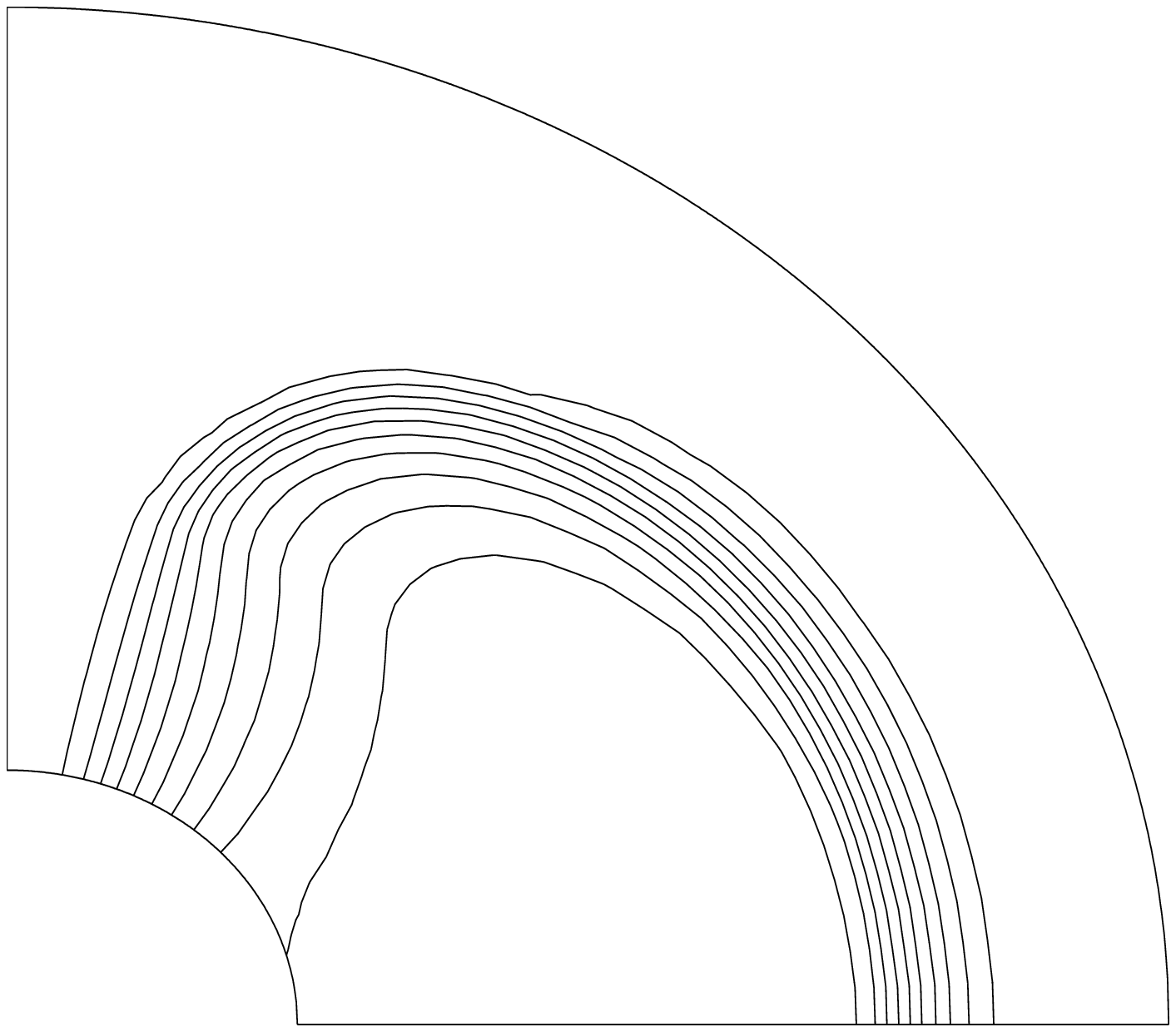}
\includegraphics[width=160pt]{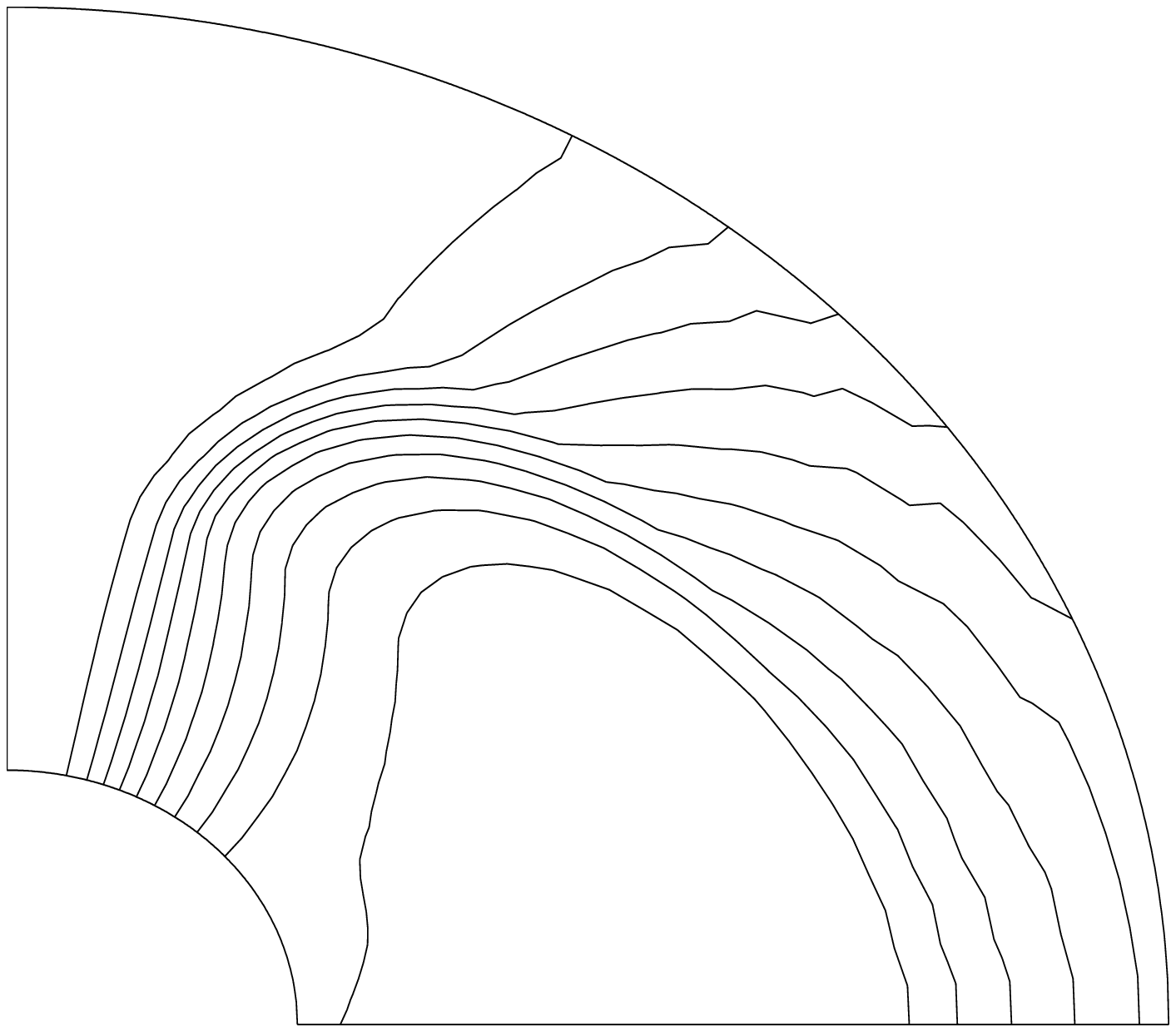}
%
\caption[]{The initial configuration of  crustal field in an idealised
  region  defined  by  $0.25  \leq  r/R_{\rm ns} \leq  1$  ($r$  -  radial
  coordinate,  $R_{\rm ns}$ -  radius of  the  neutron star)  and $0  \leq
  \theta \leq \pi/2$.   The second and the third  panel correspond to
  the field configuration  at a later time without and with magnetic
  buoyancy. See \citeN{choud02} and \citeN{konar04b} for details.}
\label{f_screen}
\end{figure*}

{\bf   Diamagnetic  screening   :}   A  number   of  mechanisms,   for
accretion-induced  field  decay, has  been  suggested  over the  years
amongst which  this probably is  the least understood process.   In an
accreting  neutron star,  material  is channelled  along the  magnetic
field lines from  the accretion disc forming accretion  columns on the
magnetic poles \cite{ghosh78,basko76}.  The  matter accumulated at the
base  of such  columns can  significantly distort  the local  magnetic
field \cite{melat01,payne04,mukhe12}.  In  particular, the field could
get    buried     under    a     mountain    of     accreted    plasma
\cite{bisno74,bland79,woosl82,hameu83,roman90,brown98a}.    When   the
amount of  accreted matter is so  large that the material  pressure at
the bottom of  the column exceeds the magnetic  pressure, the mountain
spreads laterally,  transporting the  polar magnetic flux  towards the
equator   and    finally   dissipating   them   below    the   surface
\cite{cummi01,melat01,choud02,konar04b,payne04,payne07}.

When the  accreting material  flows laterally from  the polar  caps to
lower  latitudes,  magnetic  field  lines are  dragged  by  the  flow,
creating horizontal  field at the  expense of the vertical  ones. Such
horizontal  magnetic  fields  are  known to  be  subject  to  magnetic
buoyancy  and may  cause  the field  to  pop back  up  to the  surface
\cite{sprui83} (see  Fig.\ref{f_screen} illustrating  this behaviour).
Because of  this, there is no  clear consensus on whether  large scale
field burial is possible without re-emergence \cite{cummi01} or if MHD
instabilities  would inhibit  deformation of  field lines  required to
reduce the apparent dipole moment \cite{mukhe13a,mukhe13b}.

{\bf Ohmic dissipation :}

As  mentioned before,  the  micro-physics of  field evolution  depends
crucially on the nature of the currents supporting the magnetic field.
It  is a  case of  straightforward Ohmic  dissipation in  an accretion
heated crust, if the field  is supported by crustal currents. However,
if the magnetic field exists in  the form of Abrikosov fluxoids in the
super-fluid regions of the core, then  a prior phase of flux expulsion
(to the metallic crust) would be required.

\bef 
\epsfig{file=fig03.ps,width=165pt,angle=-90}
%
\caption[]{Evolution  of  the surface  magnetic  field  strength of  a
  neutron star  as a  result of mass  accretion, assuming  the crustal
  currents to be concentrated at  a density of $10^{13}$~\gcc.  Curves
  1 to  5 correspond  to mass  accretion rates  of $\mdot  = 10^{-12},
  10^{-11},  10^{-10},  10^{-9}, 10^{-8}$~\msun/yr  respectively.  See
  \citeN{konar97a} for details.}
\label{f_ohm01} 
\eef

In an accretion-heated  crust, the decay takes place  principally as a
result  of rapid  dissipation of  currents due  to a  decrease in  the
electrical conductivity and a resultant reduction in Ohmic dissipation
time  scale \cite{geppe94,urpin95b,urpin96,konar97a,konar99a,cummi04}.
Ohmic decay  of current loops  for a  permanent decrease in  the field
strength happens according to the induction equation, given by -
\beq
\frac{\pa \bb}{\pa t}
= \nabla \times (\vb \times \bb)
  - \frac{c^2}{4\pi}  \nabla  \times (\frac{1}{\sigma}  \nabla  \times
  \bb).
\eeq
In deeper layers of the crust,  field decay is governed essentially by
the electrical conductivity $\sigma$, and the radially inward material
velocity $\vb \, (\propto \mdot/r^2$).  In turn, $\sigma$ is dependent
on  $\rho_c$,  the  density  at  which  current  carrying  layers  are
concentrated,  the impurity  content $Q$  and the  temperature of  the
crust $T_c$ (again decided by the mass accretion rate \mdot).

\bef 
\epsfig{file=fig04.ps,width=175pt,angle=-90}
%
\caption[]{The ratio  of the final  to initial surface  field obtained
  for given  values of accretion  rate in \msun/yr.  Curves  marked by
  $\rho_1,  ...,\rho_5$   respectively  correspond  to   densities  of
  $10^{11},10^{11.5},10^{12},10^{12.5},10^{13}$~\gcc,     where    the
  currents  supporting  the  magnetic   field  are  concentrated.  See
  \citeN{konar97a} for details.}
\label{f_ohm02} 
\eef

Accretion-induced heating reduces $\sigma$  and consequently the Ohmic
decay time-scale inducing  a faster decay.  At the  same time material
movement, caused  by the  deposition of  matter on  top of  the crust,
pushes the  original current  carrying layers  into deeper  and denser
regions where  higher conductivity  slows the decay  down.  Ultimately
the decay stops altogether when the original crust is assimilated into
regions  with effectively  infinite conductivity  (Fig.\ref{f_ohm01}).
Therefore, the  final saturation  field of  an accreting  neutron star
depends entirely upon the initial magnetic field, the initial $\rho_c$
and    $\mdot$    which    determine     both    $T_c$    and    {\vb}
(Fig.\ref{f_ohm02}).  Once the  surface field  freezes into  its final
value     the    equilibrium     spin-period     is    obtained     by
\cite{alpar82,chen93a}:
\ber
P_{\rm eq}  
&\simeq&  1.9 \,  {\rm ms}  
          \left(\frac{B_{final}}{10^9~{\rm G}}\right)^{6/7} 
          \left(\frac{M_{ns}}{1.4  \msun}\right)^{-5/7}  \nonumber \\
&&        \left(\frac{\mdot}{\mdot_{\rm Ed}}\right)^{-3/7} 
          \left(\frac{R_{ns}}{10^6~{\rm cm}}\right)^{16/7} \ .
\eer
where $\mdot_{\rm Ed}$ is the Eddington accretion rate.

The  intrinsic  uncertainties  associated  with  the  model  of  Ohmic
dissipation are  - a) the  impurity content of  the crust, and  b) the
exact location  of the current  carrying layers.  In  accretion heated
crust,  the effect  of impurities  can be  entirely neglected  and the
Ohmic dissipation model  can be used to constrain the  location of the
current  carrying  layers inside  a  neutron  star. Assuming  ordinary
neutron  stars  to be  born  with  magnetic  fields  in the  range  of
$10^{11.5}  - 10^{13.5}$~G,  we find  that  in order  to generate  the
observed  population of  MSRPs, the  original current  carrying layers
need to be concentrated at  $\rho_c \gsim 10^{13}$~\gcc.  Once this is
done, the constraint on the impurity content can be found.  It appears
that an impurity  concentration of $\sim 0.01$ is  consistent with the
fact that  the magnetic fields  of the  isolated radio pulsars  do not
decay    significantly    over     a    time-scale    of    $10^6$~yrs
\cite{bhatt92,hartm97}.

\bef
\epsfig{file=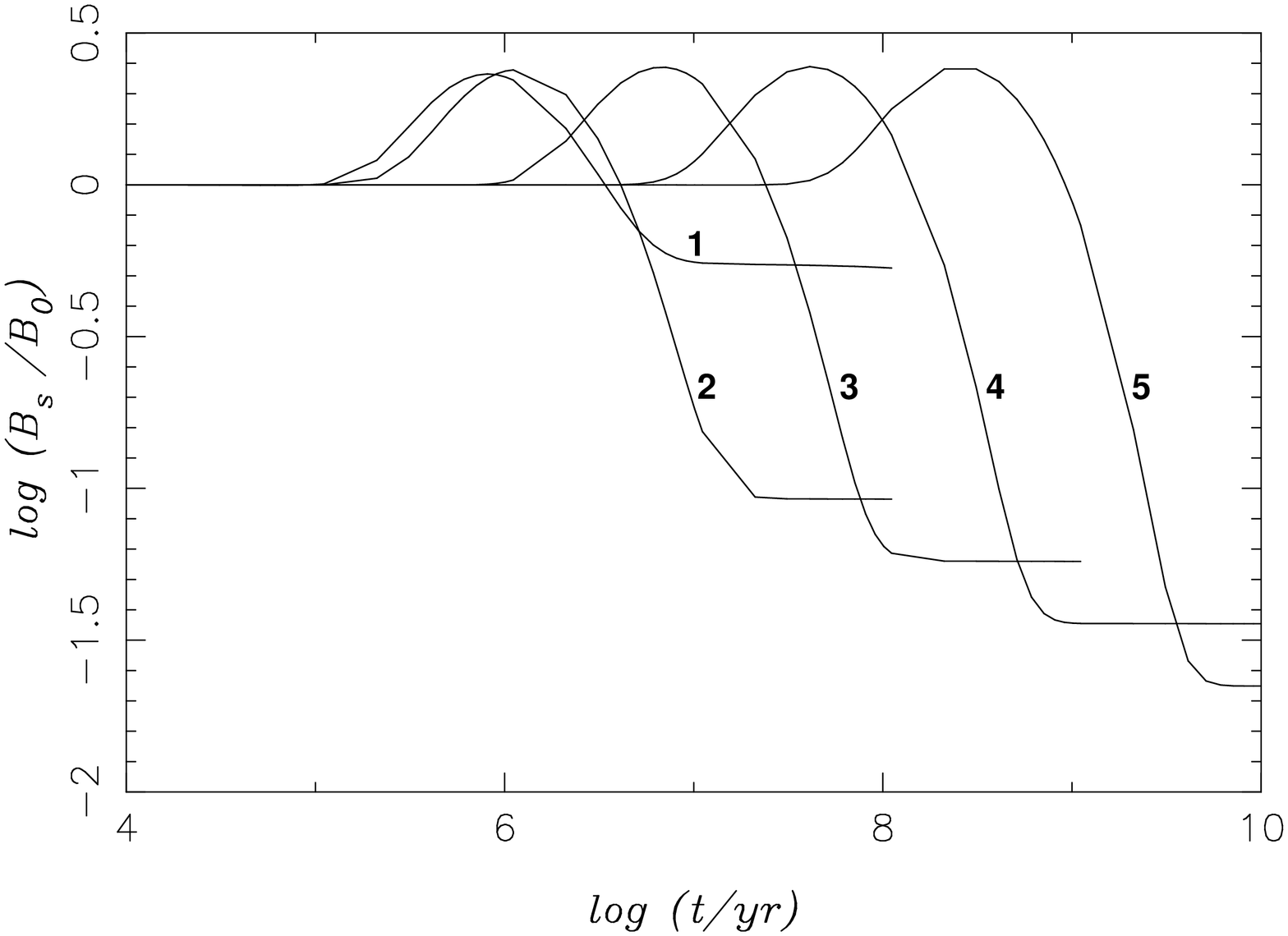,width=185pt,angle=-90}
\caption[]{Evolution  of the  surface magnetic  field for  an expelled
  flux. The  curves 1 to 5  correspond to $\mdot =  10^{-9}, 10^{-10},
  10^{-11},   10^{-12},  10^{-13}$~\dmdt.   See  \citeN{konar99b}   for
  details.}
\label{f_flux}
\eef

{\bf Spin-down induced flux expulsion :}

Inside  the core  of  a neutron  star, the  rotation  is supported  by
creation  of  Onsager-Feynman  vortices  in  the  neutron  super-fluid
whereas the  magnetic flux is  sustained by Abrikosov fluxoids  in the
proton superconductor \cite{baym69a,ruder72}.  There is likely to be a
strong  inter-pinning  between the  proton  fluxoids  and the  neutron
vortices \cite{musli85a,sauls89}.  In a spinning down neutron star the
neutron vortices migrate  outward and by virtue  of inter-pinning drag
the proton fluxoids  along to the outer crust.  A  binary neutron star
interacting  with its  companion's wind  is expected  to experience  a
major  spin-down   episode  in   the  propeller  phase,   causing  the
superconducting core  to expel a  large fraction of the  magnetic flux
\cite{srini90b}.  The  mechanism of Ohmic  decay, being unique  to the
crustal currents, is invoked for a subsequent dissipation of such flux
in the  crust \cite{jahan94,bhatt96c,konar99b,konen01a,konen01b}.  Due
to the fast dissipation of currents in an accretion heated crust, this
process may  even give  rise to  a period of  initial increase  in the
surface field  (as the core  field emerges  at the surface)  before it
starts decreasing (see Fig.\ref{f_flux}).

\subsection{Neutron Stars in HMXBs}
\label{sec03a}

\begin{table}
\begin{tabular}{|l|l|c|} \hline
&& \\
& \mdot~(\msun/year) & $\tau$~(year)\\
&& \\
  Wind Accretion
& {\color{blue}{$10^{-14} - 10^{-10}$}}  & {\color{magenta}{$10^5 - 10^7$}} \\
&& \\
  Roche Contact
& {\color{blue}{$10^{-9} - 10^{-8}$}} & {\color{magenta}{$10^3 - 10^5$}} \\
&& \\ \hline
\end{tabular}
\caption[]{Average  rates of  mass  accretion (\mdot)  by the  neutron
  stars and the associated timescales ($\tau$) during different phases
  of active mass transfer in typical HMXBs.}
\label{t_hmxb01}
\end{table}
\bef
\epsfig{file=fig06.ps,width=160pt,angle=-90}
\caption[]{The recycled pulsars from HMXBs  are expected to lie within
  the  hatched region  of this  diagram, assuming  the initial  surface
  magnetic field,  B$_{\rm s}$, to  be $\gsim 10^{12}$~G.   The dashed
  lines are  indicative of maximum  achievable spin-up, the  upper and
  lower  lines  corresponding  to  accreted masses  of  $10^{-4}$  and
  10$^{-3}$~\msun. For details please see \citeN{konar99a}.}
\label{f_hmxb}
\eef
\begin{table}
\begin{tabular}{|l|l|} \hline
  & \\
$\Delta$M~(\msun) & \color{magenta}{$10^{-6}  - 10^{-3}$} \\
  & \\  
$\Delta$L~(erg.s) & \color{magenta}{$10^{45}  - 10^{48}$} \\
  & \\  
B$_{\rm i}$~(G) & \color{blue}{$10^{11.5} - 10^{13.5}$} \\
  & \\  
B$_{\rm f}$~(G) & \color{blue}{$10^{8.5}  - 10^{13.0}$} \\
  & \\  
P$_{\rm f}$~(s) & \color{blue}{$\sim$ 0.1 - 10.0} \\
  & \\ \hline
\end{tabular}
\caption[]{The total  mass accreted ($\Delta$M) and  the total angular
  momentum ($\Delta$L) acquired by a neutron  star in an HMXB that can
  be  obtained by  assuming the  $\mdot$  and $\tau$  values given  in
  Table.\ref{t_hmxb01}.    The  corresponding   final  surface   field
  (B$_{\rm  f}$) and  spin-period (P$_{\rm  f}$) have  been calculated
  assuming the initial field value (B$_{\rm i}$) to lie in the typical
  range appropriate for ordinary non-recycled radio pulsars. }
\label{t_hmxb02}
\end{table}

The theory of the evolution of  magnetic field has received quite some
attention for neutron stars residing in  LMXBs, because of its role in
MSRP  generation.   Unfortunately,  similar  investigations  for  HMXB
systems are  yet to  be undertaken. Neutron  stars in  HMXBs typically
have B$_{\rm  s} \sim 10^{12}$~G and  O or B type  companions.  Since,
the timescales in such massive binaries are much smaller than those in
LMXBs or  IMXBs, the modification of  the magnetic field and  the spin
period would not always  be as large as that seen  in the LMXBs. Given
the average mass accretion rates  and timescales of typical HMXBs (see
Table.~\ref{t_hmxb01}) the final  values of the surface  field and the
spin-period, obtained from an Ohmic dissipation model, would typically
be  mid-range  between  the  ordinary   pulsars  and  the  MSRPs  (see
Table.~\ref{t_hmxb02}). In particular, even  when the surface magnetic
field is reduced  substantially, there may not be enough  time for the
pulsar to reach the equilibrium  spin-period. Therefore, the HMXBs are
likely to  produce low-field pulsars with  large spin-periods, perhaps
close to  the death line, as  has been shown in  Fig.\ref{f_hmxb} (for
details  see \citeN{konar99a}).   This  can also  be  understood in  a
different way.  A Gaussian decomposition of  the spin-period histogram
of  all  known  radio  pulsars   is  shown  in  Fig.\ref{f_gmm}.   The
populations  of ordinary  radio pulsars  and the  MSRPs clearly  stand
out. But, it  appears that there could be one  or more smaller clearly
identifiable  populations hidden  inside.  The  probability of  having
just one more population, midway  between the ordinary pulsars and the
MSRPs, appears to be maximum.   The location of this midway population
strongly suggests that  some of the pulsars `recycled'  in HMXBs would
actually  be  injected  into  the main  island  of  ordinary  pulsars.
Interestingly, a similar conclusion was arrived at by \citeN{deshp95a}
and \citeN{deshp95b} where  they considered the current  of pulsars in
the B$_{\rm s}$--P$_{\rm s}$ plane.

\setcounter{figure}{6}
\begin{figure*}
   \begin{center}
\epsfig{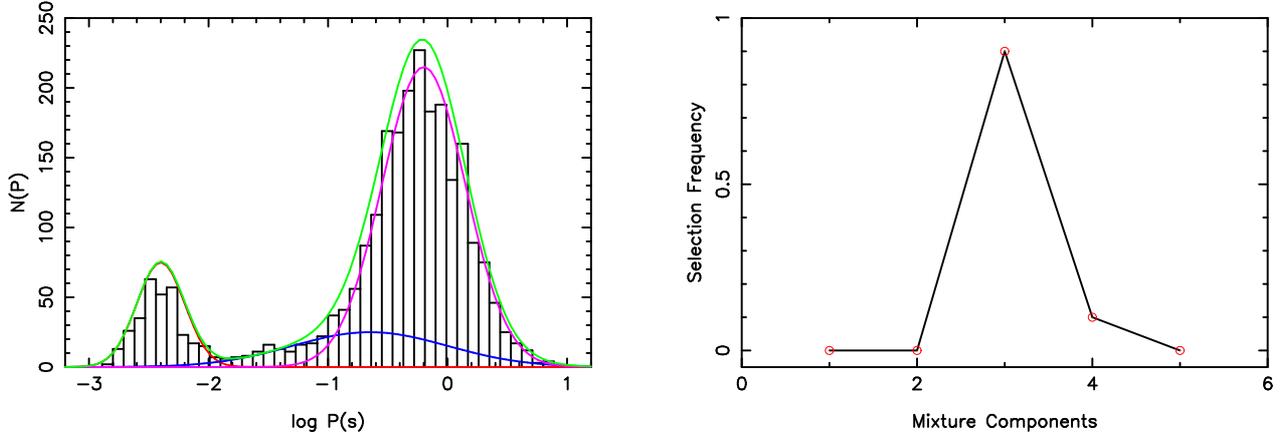}
   \end{center}
%
\caption[]{Histogram  of  the  spin-period   (P$_{\rm  s}$)  of  known
  galactic radio  pulsars.  A  Gaussian decomposition method  has been
  used to  find the  existence of uniquely  different classes  in this
  population.  The  left-hand panel shows decomposition  of the entire
  population  into three  Gaussian components.   The right-hand  panel
  shows the maximum  likelihood for a particular  decomposition with a
  given  number  of  components.  It is  evident  that  a  3-component
  decomposition is the  likeliest of all. The data is  taken from {\tt
    http://www.atnf.csiro.au/research/pulsar/psrcat/}.}
\label{f_gmm}
\end{figure*}

\subsection{Neutron Stars in LMXBs, Millisecond Pulsars}
\label{sec03b}

MSRPs are understood to attain their  fast spin and low magnetic field
strength through accretion in  LMXBs \cite{alpar82,radha82}. These are
often   found   in   binaries  with   evolved,   low-mass   companions
\cite{bhatt91a},  which are  thought to  be the  descendants of  LMXBs
\cite{helfa83,savon83,paczy83,joss83}.  This  process of  spin-up with
attendant  accretion-induced field  decay of  ordinary pulsars  to the
MSRP class is  known as `recycling'.  Therefore,  a typical progenitor
system of binary MSRPs is expected to be a neutron star accreting from
a  companion.  The  discovery of  accreting millisecond  X-ray pulsars
\cite{wijna98} and transitional millisecond pulsars \cite{papit13} has
provided confirmation to this evolutionary picture.
 

The  term  {\em millisecond  pulsar}  has  usually been  reserved  for
`recycled' pulsars with ultra-fast rotation  (P$_{\rm s} \lsim 30$ ms)
and a  weak magnetic  field (B$_{\rm  s} \sim 10^8  - 10^9$  G).  This
definition primarily uses a condition on P$_{\rm s}$.  Though the most
reasonable criterion to  define an MSRP should be to  use both P$_{\rm
  s}$  and $\pdot$  since the  evolution of  these two  parameters are
intertwined in the  formation process of MSRPs. An effort  was made to
to use  such a relation for  the MSRPs in the  Galactic disc, assuming
them   to  be   recycled   through   straightforward  LMXB   evolution
\cite{story07}. But that is not true of all MSRPs.  Moreover, for many
MSRPs  (in   particular  for  those  in   globular  clusters)  $\pdot$
determination suffers from proper  motion contamination and the values
may not be entirely reliable even when measurements are available.

On the other hand, there is some justification to use the classical $P
\ \lsim  \ 20 -  30$~ms condition  to define a  MSRP. Fig.\ref{f_gmm}
shows  a histogram  of the  spin-periods of  all known  galactic radio
pulsars. As explained in  Sec.\ref{sec03a}, the probability density of
the spin-periods is decomposed using  a Gaussian mixture model and the
data  is  best  fit  by  a  combination  of  three  distinct  Gaussian
components.   The  left-most  Gaussian  is identified  with  the  MSRP
population as  the overall probability  density shows a  prominent dip
around $P \sim 20 - 30$~ms clearly separating this population out from
the rest.

\begin{table*} 
\begin{center}
\begin{tabular}{|l|c|c|c|c|} \hline
&&&&\\ 
            & $<$ P$_{\rm s}>$ & $<$ B$_{\rm s}>$    & $<$ P$_{\rm o}>$    & $<$ M$_{\rm c}>$ \\ 
            & ms         & $10^8$~G     &               &  \msun    \\ 
MSRP &&&&\\ 
Galactic Disc :  &&&& \\  
isolated    & 5.99  & 4.64 & ------ & ------  \\
binary      & 7.68  & 8.67 & 37.10~d & 0.28  \\
BW + RB     & 2.90  & 1.80 & 7.09~hr & 0.081 \\
Globular  Clusters :  &&&& \\
isolated    & 5.89  & 13.62 & ------ & ------ \\
binary      & 5.35  & 12.63 & 8.67~d & 0.18  \\
&&&&\\ 
AMXP + AMXB & 2.77  & 0.27$^*$ & 4.61~hr & 0.14 \\
&&&&\\ \hline
\end{tabular}
\end{center}
\caption[]{A  summary  of some  of  the  average properties  of  known
  millisecond pulsars.  The  data for the MSRP is taken  from the ATNF
  online  pulsar catalogue  \cite{manch05a}.   For  AMXPs, the  magnetic
  field average has only been calculated  as the mean of the estimated
  lower limits \cite{mukhe15}.}
\label{t_msp}
\end{table*}

In  recent years,  we have  seen  millisecond pulsars  appearing in  a
variety  of types.  Like  in  the case  of  the  overall neutron  star
population, it is  likely that these different  types represent either
different phases or different  pathways of evolution.  The millisecond
pulsar types that show  distinctly different characteristics from each
other  are -  ordinary MSRP,  BW \&  RB, AMXP,  and perhaps  the newly
emerging class of Gamma-ray pulsars (which may or may not have radio
counterparts).

The AMXPs have, of course,  been identified as immediate precursors to
the MSRPs  and are expected to  move to the radio  emitting phase upon
cessation of active  material accretion. This conjecture  has now been
firmly  established with  the  detection of  three transient  pulsars,
PSR~J1023+0038,  PSR~J1227-4853  and  PSR~J1824-2452I.  All  of  these
objects alternate between a non-accreting  radio emitting phase and an
accreting X-ray phase \cite{archi09,roy15,papit13}.

The population  of AMXPs has  been growing  rapidly over the  last few
years, taking the count to 35  (inclusive of standard AMXPs and AMXBs)
\cite{mukhe15,konar17f}.    These   objects    typically   belong   to
ultra-compact  binaries,  with highly  evolved  white  or brown  dwarf
companions \cite{deloy03}, undergoing mass  transfer at very low rates
of accretion  from their low-mass  companions.  The typical  values of
\mdot~observed   in   AMXP/AMXBs  is   in   the   range  $10^{-10}   -
10^{-12}$\msun/yr.  It is understood  that objects like SAX~J1808 (the
first AMXP to  be discovered) are progenitors of fast  MSRPs with very
short orbital  periods (or  end up  isolated, given  the BW  nature of
SAX~J1808) which have  undergone very long periods  (Gyr) of accretion
at very low rates (\mdot $\sim 10^{-11}$\msun/yr) \cite{bilds01}.

The AMXPs are primarily detected  by their coherent or nearly coherent
pulsations  in persistent  X-ray  emission.   The main  characteristic
traits of these pulsars are the following.
\bei
   \i  They are  components of  transient ultra-compact  binaries with
   P$_{\rm orb} \sim$ 40~m - 20~hrs.
   \i They  usually have  ultra low-mass  companions with  M$_c \lsim$
   0.1~\msun (mostly).
   \i There appears to  be an absence of type-I X-ray  bursts and H \&
   He lines in outburst of ultra-compact  (P$_{\rm orb} <$ 1 hr) AMXPs
   suggesting that they have evolved dwarf companions.
   \i A number  of systems are observed to have  orbital periods below
   the minimum period for systems  with hydrogen donor (P$_{\rm orb} <
   80$ min).
   \i The inferred long-term $\mdot$  and L$_{\rm outburst}$ happen to
   be the lowest amongst all LMXB systems.
\eei

On  the other  hand,  the AMXBs  exhibit  nearly coherent  millisecond
oscillations during  thermonuclear type-I X-ray bursts.   As the X-ray
burst  evolves,  the  oscillation frequency  typically  approaches  an
asymptotic value  which is  stable for  a given  source from  burst to
burst.  This asymptotic frequency is thought  to trace within a few Hz
the spin frequency of the  neutron star \cite{stroh96} as  has already
been  observed for  the  AMXPs -  SAX~J1808-3658 \cite{chakr03},
XTE~J1814-338 \cite{stroh03}, Aql~X-1 \cite{casel08} etc.

An important point about AMXPs is that a number of them, including the
famous SAX~J1808  \cite{bilds01}, show  clear black widow  traits.  In
fact,   each  of   the   three   transient  pulsars   (PSR~J1023+0038,
PSR~J1227-4853, PSR~J1824-2452I) is an RB system, with a main-sequence
star  as a  companion.   It is  understood  that these  AMXP/transient
systems displaying BW/RB behaviour would  eventually join the ranks of
the isolated, rather than binary, radio millisecond pulsars with total
evaporation of  their companions.   There is some  qualitative support
for this  expectation.  Some of  the important physical  parameters of
different  millisecond   pulsar  classes   have  been   summarised  in
Table.\ref{t_msp}.  It can be seen  that the average orbital period of
the  BW/RB radio  pulsars  is much  smaller than  that  of the  binary
MSRPs. In fact, it is much closer to the average orbital period of the
AMXPs  and indicates  that these  two classes  are expected  to follow
similar evolutionary paths.

\bef
\epsfig{file=fig08.ps,width=150pt,angle=-90}
\caption{The spin-period histograms and the best-fit curves
  for AMXPs and the isolated MSRPs.}
\label{f_phst}
\eef

However, a comparison of the spin-period distribution of the AMXPs and
the  isolated  MSRPs  reveals  that  they  are  quite  different  (see
Fig.\ref{f_phst}), with  MSRPs having much larger  average spin-period
(see Table.\ref{t_msp}). At  first glance, this may  indicate that the
observed AMXPs can  not be similar to the progenitors  of the isolated
MSRPs observed  presently.  Interestingly, this discrepancy  is easily
resolved if we assume that the  AMXPs should, in fact, be treated like
`zero-age' isolated  MSRPs. Most of  the AMXP systems are  observed to
have very  low mass-accretion rates and  are likely to be  just at the
end of the accretion phase (as suggested by the intermittent accretion
seen in transient pulsars). Therefore,  as soon as the accretion stops
these AMXPs would  become full-fledged `zero-age' MSRPs.  On the other
hand, once the active accretion phase stops, the magnetic field should
stop evolving  and freeze  to its final  value \cite{konar97c}.  It is
then  reasonable to  assume  that the  AMXPS  (or `zero-age'  isolated
MSRPs) to  have magnetic  field distribution similar  to that  seen in
isolated MSRPs.  Evolving such a  population for $10^9$ years  we find
that the distribution  of the final spin periods  closely resemble the
spin    period   distribution    of    the    isolated   MSRPs    (see
Fig.\ref{f_phstmc}),  confirming  that  the AMXPs  are  indeed  direct
progenitors of the isolated MSRPs.

\bef
\epsfig{file=fig09.ps,width=150pt,angle=-90}
\caption[]{A simulated population of  zero-age isolated MSRPs with the
  spin-distribution  of AMXPs  (red histogram)  is evolved  for $10^9$
  years assuming the magnetic field  remains constant. It evolves into
  a population (blue histogram) that closely resemble the present day
  isolated MSRP population.}
\label{f_phstmc} 
\eef

A large number  of pulsars has been detected in  the globular clusters
in  recent  years and  they  appear  to have  somewhat  different
characteristics than  the pulsars observed  in the Galactic  disc (see
Table.\ref{t_msp}).   To begin  with,  it is  found  that the  average
orbital  period of  the  binary millisecond  pulsars  in the  globular
clusters is neither similar to that  in the disc nor to the AMXP-BW/RB
set. This  itself is an  indication that  the binary evolution  in the
clusters  are  probably  different.  The conditions  prevailing  in  a
globular cluster are rather different  from those in the Galactic disc
primarily due to the extremely high stellar densities in the clusters.
One of the obvious effects of this high density is a dramatic increase
in the  number of binaries,  as well as in  the rate of  close stellar
encounters allowing for many  different channels for binary formation.
As  a  consequence,  most  binaries are  not  primordial  in  globular
clusters.  For example, the total observed number of LMXBs in globular
clusters exceeds their formation rate in the disc by several orders of
magnitude,  indicating  a  dynamical origin  \cite{clark75}.   Stellar
interactions  in  the clusters  involving  a  neutron star  have  been
studied         by         a         number         of         authors
\cite{kroli84,rasio91,davie92,davie98,rasio00b,bagch09a,bagch09b}.
Moreover,  many of  the cluster  binaries have  properties similar  to
those of the black widow pulsars  seen in the Galactic disk population
\cite{king03b,freir05a}.

It is also seen that a subset  of slower MSRPs in the globular clusters
(for which  field measurements are  available) appear to  have surface
magnetic  fields that  are 2-5  times  larger compared  to their  disc
counterparts;  even though the spin-periods  of this subset of cluster
pulsars  are  similar   to  the  spin-periods  of  the   MSRPs  in  the
disc. Likely,  the difference in  the surface magnetic field  could be
due   to  one   or   several  of   the   possibilities  listed   below
\cite{konar10}:
\bei
\i There are  systematic biases in the \pdot\  (hence B$_{\rm s}$) measurement
for cluster MSRPs.
\i  The cluster  MSRPs are  younger and  may evolve  to  a distribution
similar to the disc MSRPs with time.
\i Preferential recycling of MSRPs  in tighter binaries with high rates
of  attendant mass  transfer may  actually result  in  cluster pulsars
retaining higher magnetic fields.
\eei

\section{Magnetic Field : Isolated Evolution}
\label{sec04}

The basic  physics, underlying the model  of magneto-thermal evolution
developed  for the  evolution of  magnetic field  in isolated  neutron
stars, is  essentially the same as  in the case of  accreting systems.
However, in absence of any  external factors (like material accretion)
the evolution feeds upon itself.  On the one hand, the field evolution
is  sensitively  dependent  on  the  micro-physics  (through  transport
properties)  of  the  interior.  Since the  transport  properties  are
temperature-dependent, thermal evolution  affects field evolution.  On
the other  hand, the  Ohmic dissipation of  the field  generates heat,
modifying  thermal   evolution,  changing  transport   properties  and
ultimately affecting the field evolution in a cyclical fashion.

In  recent  years,  there  have  been  observational  indications  for
existence  of  evolutionary  pathways  linking  different  classes  of
isolated  neutron  stars \cite{kaspi10}.   There  is  a clear  overlap
between  the high  magnetic field  (B$  > 4  \times 10^{13}$~G)  radio
pulsars  and  the  magnetars  in the  B-P  diagram  (Fig.[1]).   The
magnetar-like X-ray burst  exhibited by PSR~J1846-0258 (B$  = 4 \times
10^{13}$~G) has reinforced  the suggestion that such  high field radio
pulsars are  quiescent magnetars.   Conversely, it has  been suggested
that hyper-critical  fallback accretion may  bury the field  to deeper
crustal  layers thereby  reducing the  surface field,  as seen  in the
CCOs.  Subsequent emergence of this  buried field could transform a
CCO to  an ordinary  radio pulsar  or even  to a  magnetar.  Therefore
different  combinations of  initial  spin-period,  magnetic field  and
submersion depth of  the field may very well decide  whether a neutron
star manifests itself as an ordinary radio pulsar, a magnetar or a CCO
\cite{vigan12}.  Similarly,  INSs are observed only  in X-ray, despite
being isolated objects.  It is possible that they are actually similar
to  the RPPs  and are  not seen  as radio  pulsars simply  due to  the
misalignment of emission  cones with our lines of  sight.  The neutron
stars  with  strong  magnetic  fields  are expected  to  remain  at  a
relatively higher  temperature due  to field  decay.  This  could then
explain the high (compared to ordinary radio pulsars) X-ray luminosity
of the  INSs. Finally, it has  been argued that the  anomalous braking
index of PSR~J1734-3333 signifies an increase in  its dipolar surface
magnetic  field  and  is  likely  driven  by  the  emergence  (perhaps
glitch-induced)  of a  stronger field  buried underneath  the surface,
with  timescales depending  on submersion  depth \cite{espin11c}.   If
correct,  this process  may chart  a pathway  for the  transition from
ordinary  radio  pulsars  to  magnetars.  It  appears  that  different
flavors of the  isolated neutron stars could, in  fact, be intricately
connected through various evolutionary pathways.

One  of the  important ingredients  of the  theory of  magneto-thermal
evolution, effective in the early phases  of a neutron star's life, is
the structure of the magnetic field.   Many of the models consider the
interaction of the observable poloidal field with a toroidal component
buried  in  the  deeper  layers  of the  crust.   The  magnetic  field
estimated from the  spin-down rate of radio pulsars  measures only the
large scale dipolar field. However, strong multipole components of the
magnetic field have long been thought to play an important role in the
radio               emission                from               pulsars
\cite{ruder75,kroli91,arons98,deshp98,ranki98,gil02a,gil02b,asseo02,zane05,gil08}.
In  absence of  material  movement  (as can  be  expected in  isolated
neutron stars) each multipole  component evolves separately for purely
poloidal field  structures. As  higher multipole  components dissipate
much faster than the dipole, it is expected that over long time-scales
only the dipole field would survive \cite{mitra99}. Since existence of
higher multipoles is  absolutely necessary for pulsar  activity, it is
imperative that  the internal  magnetic field of  a neutron  star must
have a toroidal component. Of course, existence of toroidal components
is also necessary for the stabilisation  of poloidal fields in a newly
born fluid neutron  star before anchoring in the solid  crust can work
towards stabilising the large scale field.

\section{Discussions}
\label{sec05}

\begin{figure*}
\begin{center}
\includegraphics[width=14.0cm]{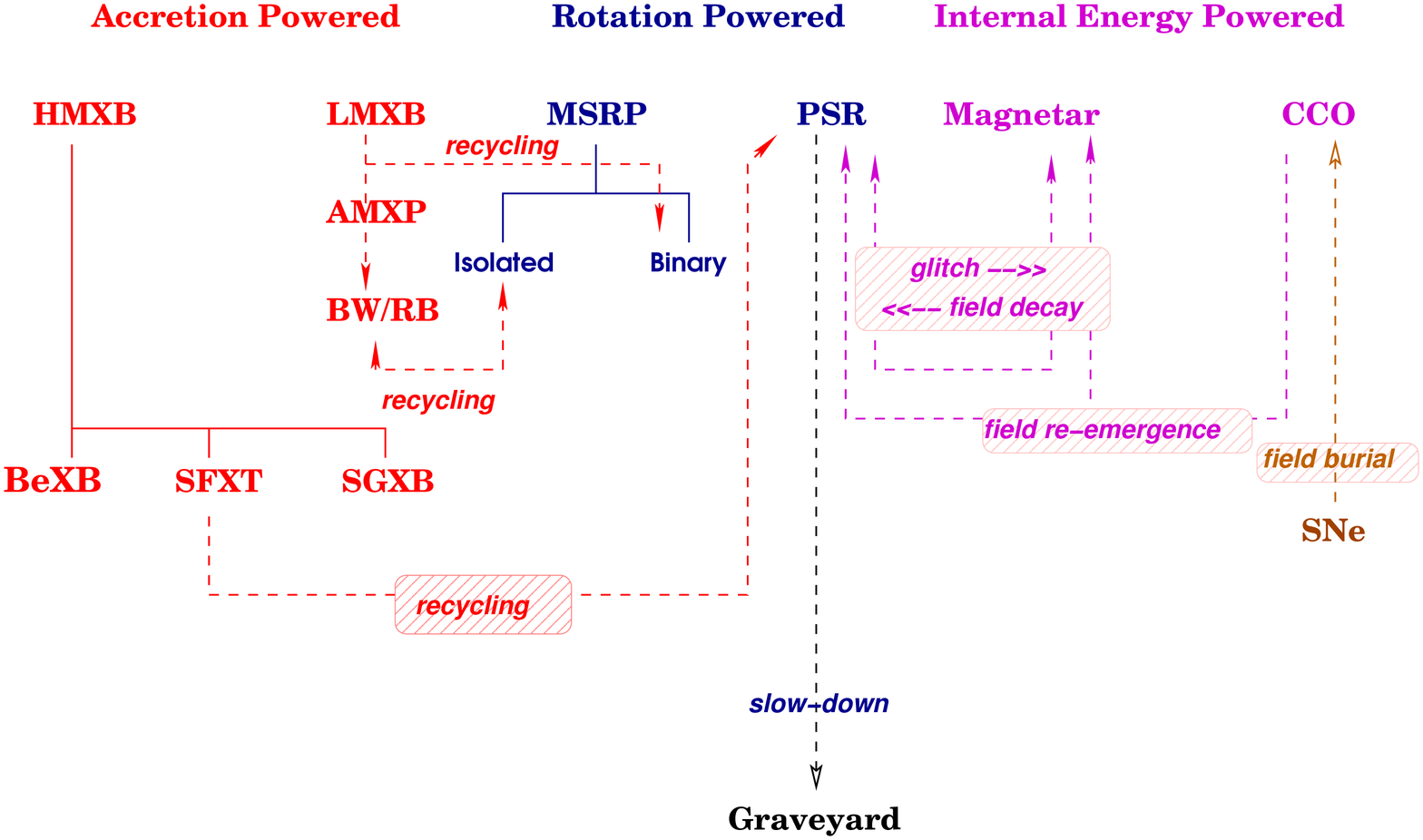}
\end{center}
\caption[]{A summary of the evolutionary pathways connecting different
  observational  classes  of neutron  stars.   Wherever  put inside  a
  shaded box, the pathway is not  a confirmed process yet.  (Legends :
  HMXB  - high  mass X-ray  binary,  BeXB -  Be X-ray  binary, SFXT  -
  super-giant  fast X-ray  transient, SGXB  - super-giant  X-ray binary,
  LMXB - low mass X-ray binary,  MSRP - millisecond radio pulsar, AMXP
  - accreting  millisecond X-ray  pulsar, BW/RB  - black-widow/redback
  pulsar, PSR  - radio  pulsar, CCO  - central  compact object,  SNe -
  supernova explosion)}
\label{f_evolution}
\end{figure*}

The evolutionary pathways, linking  different observational classes of
neutron stars,  has been  summarised in Fig.\ref{f_evolution}.   It is
evident that on the fiftieth year  of the discovery of the first radio
pulsar,  we have  unearthed more  classes than  we have  been able  to
link.  Much  of  the  pathways, in  particular  between  the  isolated
variety,  is  quite uncertain  and  is  still  being explored.  It  is
expected that the number of new neutron stars discovered will increase
by  a   large  factor  in   the  near   future  with  the   advent  of
multi-messenger   astronomy   and   future  telescopes   with   better
sensitivity  and   wider  frequency   coverage.  Such   increase  will
definitely improve  our understanding  of some of  today's `uncertain'
pathways but almost certainly will throw up newer challenges.

\section*{Acknowledgements}

The author  has been privileged  to begin  her academic career  in the
astrophysics group of Raman Research Institute, Bangalore; then headed
by G.  Srinivasan and to learn many of the intricacies of neutron star
physics from  him. Most  of the  work reported here  has been  done in
collaboration  with   Dipankar  Bhattacharya,  Arnab   Rai  Choudhuri,
Dipanjan  Mitra and  Dipanjan Mukherjee\footnote{All  of these  people
  have been  associated with the  `Srini era' of RRI  astrophysics, in
  one capacity or  another.  Except for Dipanjan  Mukherjee who worked
  with Dipankar Bhattacharya, himself a  student of Srini, for his PhD
  thesis.}.     The     author    is    supported    by     a    grant
(SR/WOS-A/PM-1038/2014) from DST, Government of India.

\bibliography{mnrasmnemonic,adsrefs}
\bibliographystyle{mnras}

\end{document}